\documentstyle{article}
\begin{document}
\setlength{\baselineskip}{15pt}
\title{Mass action law conjugate representation for general chemical 
       mechanisms}
\author{V\'{\i}ctor Fair\'{e}n$^*$ \and Benito Hern\'{a}ndez--Bermejo}
\date{}
\maketitle

{\em Departamento de F\'{\i}sica Fundamental, Universidad Nacional de 
Educaci\'{o}n a Distancia. Apartado 60.141, 28080 Madrid (Spain).
Email: vfairen@uned.es \/}
   
\begin{abstract}
Power-law rates constitute a common approximation to the general analysis 
of the stability properties of complex reaction networks. We point out in 
this paper that this form for the rates does not need to be assumed as an 
approximation for general rate-laws. On the contrary, any functional form 
for a rate law can be represented exactly in terms of power-laws. Moreover, 
we can uniquely associate to any set of kinetic equations an equivalent 
`conjugate' representation in terms of the well-known generalized 
Lotka-Volterra equations, standing for what we call {\em per capita\/} 
rates, which amounts to a great simplification in terms of the structural 
form of the mathematical representation of a reaction network. 

\end{abstract}

\mbox{}

\mbox{}

\mbox{}

\mbox{}

\mbox{}

\mbox{}

\mbox{}

\mbox{}

\mbox{}

\mbox{}

\mbox{}

\mbox{}

\mbox{}

\mbox{}

$^{*}$ Author tho whom all correspondence should be addressed.

\pagebreak
\begin{flushleft}
{\bf I. Introduction}
\end{flushleft}

Mass action law chemical kinetics, and its corresponding mathematical 
modelling, has for long been considered as a prototype in nonlinear 
science.$^1$ We can all recall how such archetypical schemes, as the 
Brusselator, Oregonator, Schl\"{o}gl model (to cite a few) which have 
constituted the vanguard in the pioneering years. The simplicity of the 
stoichiometric rules and that of the algebraic structure of the corresponding 
evolution equations has made chemical kinetics a traditional point of 
reference in modeling within such fields as population biology,$^2$ 
quantitative sociology,$^3$ prebiotic evolution$^4$ and other biomathematic 
problems,$^5$ where a system is viewed as a collection of `species' 
interacting as molecules do. Moreover, as emphazised by \'{E}rdi and 
T\'oth,$^1$ even the algebraic structure of the evolution equations from many 
other fields can be converted into  `chemical language', where a formal 
`analog' in terms of a chemical reaction network is defined. 

The interest of this common mathematical framework provided by chemical 
kinetics is not only aesthetic. It has sparked the quest of theorems which 
connect the structure of the chemical reaction network with the qualitative 
features of the solutions to the corresponding differential equations.$^6$ 
In fact, we are talking about the search of theorems which would permit the 
knowledge of behaviors open to the system from an identification of certain 
patterns in the network, and that of the associated algebraic structure of 
the differential equations. And conversely, mathematical propositions which 
should eventually point at which of the properties of a chemical reaction 
network are to be selected for the obtainment of a given behavior. The 
ultimate goal being that of a classification of networks, or at least of 
certain of their characteristics. The accomplishment of this purpose would  
certainly yield a tool of great practical importance in modeling. The 
{\em zero deficiency theorem\/}$^{7,8}$ and {\em Vol'pert's theorem\/}$^9$ 
are examples of results associating graph properties of the reaction network 
with the existence of equilibrium points. 

In the context of classification of networks, a significant step forward has 
been done by Clarke$^{10}$ with the stoichiometric network analysis, with 
the help of which he addresses the issue of connecting the topology of a 
given chemical network with: 1) {\em The network stability problem\/} 
(necessary and sufficient conditions for ensuring stability of steady states); 
2) {\em The stability diagram problem\/} (calculation of the bifurcation set 
of an unstable network). Ross  and collaborators$^{11}$ have made extensive 
use of this tool for the categorization and obtainment of model mechanisms 
from experimental data in chemical oscillators. 

As emphasized by Clarke, the key role in the stoichiometric network analysis 
is played by the effective power function for species $i$ in reaction $j$, 
defined at steady state ${\bf X}_0$ as:
$$    
    \kappa _{ij}({\bf k}) = \left[
      \frac{ \partial \log v_j ({\bf X},{\bf k})}{ \partial \log X_i } 
      \right] _{{\bf X} = {\bf X}_0} \;\: ,   \eqno{(1)}
$$    
where $ v_j ({\bf X},{\bf k}) $ is the reaction rate, dependent upon a 
concentration vector {\bf X}, and a set of reaction constants, {\bf k}. Then, 
the stability of the steady state, in a network involving $n$ species and 
$r$ reactions, is given by the solutions of:
$$    
   \frac{ \mbox{d}}{ \mbox{d} t} \; \delta X_i \; = \sum_{m=1}^{n} \left[ 
       \sum_{j=1}^{r} \nu _{ij} \kappa _{mj} \left[ \frac{v_j}{X_m} \right] 
       _{{\bf X} = {\bf X}_0} \right] \delta X_m      \eqno{(2)}
$$    
where $\nu _{ij}$ are the elements of the net stoichiometric matrix.

In principle, the rate laws $ v_j ({\bf X},{\bf k}) $ can have any form, 
but in (1) and (2) we effectively assume that in practice rate 
laws can be approximated locally, around the steady state, by the traditional 
power-law,
$$    
      v_j ({\bf X},{\bf k}) = k_j \prod_{i=1}^{n} X_i^{\kappa _{ij}} \;\: ,
      \eqno{(3)}
$$    
in which the reaction orders $\kappa _{ij}$ are not restricted to to the integer 
values given by the law of mass action.  These non integer values have 
already been found in situations where the mean-field approximation does 
not hold,$^{12}$ as it happens in diffusion-limited kinetics, either in 
disordered media with fractal structures, or on regular lattices of 
dimensionality smaller than the critical value, $d=2$.$^{13}$ Here, the 
single elementary reaction 
\[
      2A \longrightarrow \mbox{products}
\]
displays a non-integer order of reaction, which may be even greater than 3 
when occuring on fractal `dust',$^{14}$ with spectral dimension $0 \leq d_s 
\leq 1$. Also, an extensive use of the power-law approximation has been made 
by Savageau and collaborators. Starting with the observation that 
enzyme-kinetic rates are well represented by linear relations in logarithmic 
space,$^{15}$ they have generalized this structural pattern to the analysis 
of many natural systems, encapsulating their modeling in a systematic use 
of a version of the power-law formalism called S-system approach.$^{16}$ 
Finally, we can mention the different solutions suggested to solve the 
`inversion problem': the embedding of general differential equations into a 
unified formalism in terms of stoichiometric networks with power-law 
kinetics; or, more specifically, mass-action kinetics.$^{17}$ In this 
context, contributions from Samardzija {\em et al.\/},$^{18}$ Poland$^{19}$ 
and Kowalsky$^{20}$ have tried different routes for producing stoichiometric 
network counterparts of well known prototypical models, as Lorenz and 
R\"{o}ssler systems, or the Van der Pol oscillator.

The assumption of power-law rates is at the heart of most treatments trying 
to establish a unifying mathematical framework around the concept of a 
stoichiometric network. This systematic approach opens, as Ross and 
collaborators have shown,$^{11}$ new horizons to the chemical dynamicists, 
inasmuch as the structural analogies which might be discovered will be of 
help in configuring an association between the structure of a chemical 
mechanism and the expected behavior. 

We intend in the present article to stress how general functional forms for 
the rate laws may be exactly encapsulated into a power-law formalism without 
resorting to a local approximation, as in (3). We will then rewrite the 
resulting kinetic equations into power-law rates equations, and that will 
permit to show that the evolution equations for the {\em per capita\/} 
rates, defined as
\[
   \frac{ v_j ({\bf X},{\bf k}) }{ X_i } \;\: ,
\]
are always in the form of generalized Lotka-Volterra equations, no matter 
what is the particular form of the original rate equations. An equivalent 
`conjugate' network may be associated to this generalized Lotka-Volterra 
representation, which involves unimolecular, bimolecular and 
pseudounimolecular steps. We shall discuss the properties of this 
transformation and show that it leads to a unique generalized Lotka-Volterra 
representation for a given reaction network. 

\mbox{}

\begin{flushleft}
{\bf II. Exact equivalence to power-law rates systems}
\end{flushleft}

Within the power-law formalism, the kinetic equations for a given species 
involved in a mechanism with $r$ reactions are:
$$    
    \dot{X}_i = \sum_{j=1}^{r} k_j \nu _{ij} \prod_{k=1}^{n} X_k^{ 
    \kappa _{kj}} \;\: ,    \eqno{(4)}
$$    
to which we shall refer, from now on, as power-law rates systems.

The question now is to demonstrate that general functional relations for the 
rate laws are amenable to an equivalent power-law form without resorting to 
approximations. The procedure to do so is well known$^{21,22}$ and can be 
illustrated with a simple example. Assume the following mechanism of 
pseudoreactions
$$   
     A \stackrel{k_1}{\longrightarrow} Y    \eqno{(5.a)}
$$   
$$   
     B \stackrel{k_2}{\longrightarrow} X    \eqno{(5.b)} 
$$   
$$   
     X \stackrel{k'_2}{\longrightarrow} B   \eqno{(5.c)} 
$$   
$$   
     X + Y \stackrel{k_3}{\longrightarrow} \mbox{products}  \eqno{(5.d)} 
$$   
which constitute an early model by Degn and Harrison$^{23,24}$ to account for 
the oscillations in the peroxidase-oxidase reaction:
\begin{center}
    2 NADH + O$_2$ + 2H$^+$ $\longrightarrow$ 2 NAD$^+$ + 2H$_2$O 
\end{center}
Reaction (5.d) is the peroxidase enzyme catalyzed oxidation of the 
NADH ($Y$) by dilute oxygen ($X$), which was assumed in the model to be 
inhibited at high concentrations of the latter. According to Degn and 
Harrison the corresponding rate law was taken to follow a Michaelis-Menten 
form, suggesting for model (5), when species $A$ and $B$ are held constant, 
the following equations in dimensionless-form:
$$  
  \dot{X}  =  B - X - \frac{XY}{1+qX^2}  \eqno{(6.a)}
$$
$$
  \dot{Y}  =  A - \frac{XY}{1+qX^2}      \eqno{(6.b)}
$$
We now introduce the auxiliary variable $Z=(1+qX^2)^{-1}$, which converts the 
r.h.s. of (6) into polynomial form, but which calls for a supplementary 
equation for that same auxiliary variable. It will be 
$$
  \dot{Z} = \frac{ \partial Z }{ \partial X } \dot{X} \;\: ,
$$
which again is polynomial provided $ \partial Z / \partial X $ is already in 
such form. This can be proved to be the case for smooth functions, though we 
shall not discuss the demonstration here (interested readers are referred to 
Kerner$^{21}$ and Hern\'{a}ndez--Bermejo and Fair\'{e}n$^{22}$). After 
elementary algebra, we find for (6) 
$$
  \dot{X}  =  B - X - XYZ    \eqno{(7.a)}
$$              
$$
  \dot{Y}  =  A - XYZ        \eqno{(7.b)}
$$              
$$
  \dot{Z}  =  -2qBXZ^2 + 2q X^2 Z^2 + 2q X^2 Y Z^3 \;\: ,  \eqno{(7.c)}
$$
which is written in terms of power-law rates. In going from (5) to (7) the 
dimensionality of the kinetic equations has been increased. The equivalence 
between these two sets of equations will be ensured if the initial condition 
for variable $Z$ is taken to be $Z(0) = (1+qX(0)^2)^{-1}$ (we again refer to 
Kerner$^{21}$ for further details). According to (7) the step in (5.d) is 
substituted by the `kinetically equivalent' set of pseudoreactions
$$
     X + Y + Z \longrightarrow Z + \mbox{products}         \eqno{(8.a)}
$$
$$
     Q + B + X + 2Z \longrightarrow X + \mbox{products}    \eqno{(8.b)}
$$
$$
     Q + 2X + 2Z \longrightarrow 2X + 4Z                   \eqno{(8.c)}
$$
$$
     Q + 2X + Y + 3Z \longrightarrow 2X + Y + 5Z           \eqno{(8.d)}
$$
The high molecularity of the pseudoreactions (8) might eventually be reduced 
if additional `auxiliary variables' were introduced in (7), i.e.: $XZ=W$; 
with a corresponding increase in the number of items in the kinetically 
equivalent set of pseudoreactions. 

The previous procedure can be systematically carried out$^{22}$ for any 
rate-law represented by a smooth function. It amounts to a reduction in the 
degree of nonlinearity to power-laws by the labelling under `auxiliary 
variables' of functional expressions of the original independent variables. 
The initial conditions for these new variables are then automatically 
prescribed by the same functional expressions from which they are defined. 
Once the procedure is complete, the new variables are understood to represent 
the concentrations of some `virtual' species, reacting in accordance to some 
appropriate mechanism, as in (8).

Up to this point, and before proceeding any further, a comment is needed on 
the procedure leading from (5) to (8). The previous method should be viewed 
as a `protocol' for rewriting a system of ordinary nonlinear differential 
equations into a pattern formally identifiable as one describing the 
evolution of a collection of `objects' interacting according to the rules of 
the law of mass action. We then do refer to `pseudospecies', rather than to 
chemical species, because no actual chemical process has been found to obey 
such schemes (network). This is simply due to the fact that the combinations 
and scenarios open to objects behaving under the simple rules of the law of 
mass action clearly outnumber the actually known chemical processes, 
including those which are seriously considered chemically plausible but have 
not been actually obseved.

\mbox{}

\begin{flushleft}
{\bf III. Conjugate representation in terms of generalized Lotka-Volterra 
equations}
\end{flushleft}

We can now return to (4) and write it in a slightly different way:
$$
    \dot{X}_i = X_i \sum_{j=1}^{r} k_j \nu _{ij} \prod_{k=1}^{n} 
      X_k^{\kappa _{kj} - \delta _{ik}} = X_i \sum_{j=1}^{m} A_{ij} \prod_{k=1}^{n} 
      X_k^{B_{jk}} \eqno{(9)}
$$
where $i = 1, \ldots , n$ and $\delta _{ik}$ is the Kronecker delta symbol. 

The kinetic equations are now written in terms of the {\em per capita\/} 
rates 
$$    
    Y_j = \prod_{k=1}^{n} X_k^{B_{jk}} \; , \;\:\; j = 1, \ldots , m
   \eqno{(10)}
$$
In (9) we implicitly assume that there are actually $m$ distingishable 
{\em per capita\/} rates in a network with $r$ reactions: $m$ is not 
necessarily equal to $r$, for different reactions might possess the same 
{\em per capita\/} rates, and conversely, a single reaction will generate a 
specific {\em per capita\/} rate for each of the relevant species involved. 

We shall henceforth assign, in (9) and (10), the label $j=1$ to the constant 
{\em per capita\/} rate (that with $B_{1k} = 0, \;\: k = 1, \ldots , n$). It 
will then be understood that $Y_2, \ldots , Y_m$ will have at least one 
non-null $B_{jk}$ entry. Then, {\bf B} will be an $m \times n$ matrix with a 
null first row, and {\bf A} an $n \times m$ matrix with its first column 
filled in with the coefficients of $Y_1$ (which will correspond to the linear 
rates in the network). 

Let us now take time derivatives of both sides of (10). For $Y_1$ we have
$$
     \dot{Y}_1 = 0  \eqno{(11)}
$$
to which we can assign, without loss of generality, the solution
$$
     Y_1 = 1 \eqno{(12)}
$$
On the other hand, for $Y_2, \ldots , Y_m$, we have
\[
   \dot{Y}_j = \sum_{s=1}^{n} \frac{ \mbox{d} Y_j }{ \mbox{d} X_s} \dot{X}_s 
   = \sum_{s=1}^{n} B_{js} \prod_{k=1}^{n} X_k^{B_{jk} - \delta_{ks}} = 
\]
$$
   = Y_j \sum_{q=1}^{m} \left( \sum_{s=1}^{n} B_{js} A_{sq} \right) 
   \prod_{p=1}^{n} X_p^{B_{qp}} =
   Y_j \sum_{q=1}^{m} L_{jq} Y_q   \eqno{(13)}
$$
Then, the resulting equations of motion for the {\em per capita\/} rates 
are in the form of generalized Lotka-Volterra equations. They might 
themselves, in turn, be assigned to the time evolution of a set of 
pseudospecies in a `chemical network' comprising, at most, bimolecular steps 
in the relevant pseudospecies. This result is universal and applicable to 
any set of kinetic equations with power-law rates, and by extension, as 
demonstrated before, it is also valid for general functional relations for 
the rates. 

The generalized Lotka-Volterra equations (13) constitute a conjugate 
representation of the original kinetic equations (4) for a given 
chemical network. They provide a much stronger unifying structure than that 
associated to (4); a structure for which there exist several tools for 
studying the features of the solutions,$^4$ and which are straightforwardly 
translatable into a graph theory setting.$^{25}$ This is not the place to 
discourse upon the issues related to the generalized Lotka-Volterra 
equations, because most of the general results of interest to the chemical 
dynamicist (related to stability of steady states) are highly 
mathematical$^4$ and fall outside the scope of this paper. We shall 
nevertheless touch upon some aspects of interest later on, after giving an 
example of the procedure of obtainment of (13). 

{\bf Example:} Let us take (7). According to the notation in (9) we can write 
for matrices {\bf A} and {\bf B}: 
$$
{\bf A} = \left(   \begin{array}{ccccccc}
       -1 & B  &  -1  &  0  &  0    &  0  & 0     \\
        0 & 0  &   0  &  A  & -1    &  0  & 0     \\
        0 & 0  &   0  &  0  & -2qB  & 2q  & 2q 
   \end{array} \right)  \eqno{(14)}
$$
$$
{\bf B} = \left(   \begin{array}{ccc}
       0 & 0 & 0 \\ -1 & 0 & 0 \\ 0 & 1 & 1 \\ 0 & -1 & 0 \\ 1 & 0 & 1 
       \\ 2 & 0 & 1 \\ 2 & 1 & 2
       \end{array} \right) \; , \;\:  \eqno{(15)}
$$
from which we can easily calculate matrix ${\bf L} = {\bf B} \cdot {\bf A}$, 
in (13). {\bf B} is simply the order matrix for the {\em per capita\/} rates 
involved in (7), and can be systematically written down once we display 
(7) in the following form:
$$
 \frac{ \mbox{d}}{\mbox{dt}} \ln (X) = \frac{\dot{X}}{X}  =  
 - 1 + B X^{-1} - YZ    \eqno{(16.a)}
$$
$$
 \frac{ \mbox{d}}{\mbox{dt}} \ln (Y) = \frac{\dot{Y}}{Y}   =  
 A Y^{-1} - XZ          \eqno{(16.b)}
$$
$$
 \frac{ \mbox{d}}{\mbox{dt}} \ln (Z) = \frac{\dot{Z}}{Z}   =  
 -2qBXZ + 2q X^2 Z + 2q X^2 Y Z^2 \;\: ,   \eqno{(16.c)}
$$
whilst matrix {\bf A} is obtained from the corresponding coefficients. As 
indicated before, the first row in (15) has zero entries. Correspondingly, 
the generalized Lotka-Volterra matrix ${\bf L} = {\bf B} \cdot {\bf A}$ 
will also have a zero first row, which is understandably assignable to (11). 

A mechanism (or network) may be associated to the generalized Lotka-Volterra 
equations in terms of unimolecular, pseudounimolecular and bimolecular steps 
which follow the law of mass-action. This mechanism is a sort of `conjugate' 
reactional scheme to the original network, and might be as well be used as 
a complementary (or alternative) representation in investigations directed 
torwards the classification of complex reaction networks.$^{10}$ The 
structural simplicity of the generalized Lotka-Volterra equations (which are 
characterized by a single algebraic object: matrix {\bf L}, the properties of 
which are strongly related to certain important features of the solutions), 
as well as their ubiquity in many scientific disciplines,$^{26}$ makes this 
approach particularly attractive.

Upon examination of matrix {\bf L} obtained from, say, (14) and (15), we 
infer that its last three rows will be linear combinations of the three 
immediately preceding ones. By construction, this fact is generalizable to 
any matrix {\bf L} ($m-n-1$ rows will be linearly dependent on $n$ rows). 
This pattern determines the establishment of an associated {\em modular\/} 
mass-action law reaction network. A pseudoreaction {\em template\/} is 
ascribed to each entry of the independent rows in matrix {\bf L}. Those 
pseudoreactions templates constitute the modules, or building blocks, which 
the whole reaction network is made of. 

In connection to the example of the peroxidase-oxidase model (7), these 
constitutive units are, in view of (14) and (15), given by 
$$
       \pm Y_j ( -1 -BY_2 + Y_3)                     \eqno{(17.a)}
$$
$$
       \pm Y_j ( AY_4 -(1+2qB)Y_5 + 2qY_6 + 2qY_7 )  \eqno{(17.b)} 
$$
$$
       \pm Y_j ( -AY_4 + Y_5 )                       \eqno{(17.c)}  
$$
for any $j \geq 2$.

Expression (17.a) might, for example, schematically represent any of the 
following pseudoreaction templates
\[
   \begin{array}{cl}
   \mbox{with (+) sign } & 
   \left\{ \begin{array}{ccc}
        Y_j             & \longrightarrow & \mbox{}           \\
        Y_j + B + Y_2   & \longrightarrow & B + Y_2           \\
        Y_j + Y_3       & \longrightarrow & 2Y_j + Y_3
   \end{array} \right.                                              \\
    & \\
   \mbox{with (--) sign } & 
   \left\{ \begin{array}{ccc}
        Y_j             & \longrightarrow & 2Y_j              \\
        Y_j + B + Y_2   & \longrightarrow & 2Y_j + B + Y_2    \\
        Y_j + Y_3       & \longrightarrow & Y_3
   \end{array} \right.
   \end{array}
\]

\mbox{}

\begin{flushleft}
{\bf IV. Properties of the transformation to the conjugate representation}
\end{flushleft}

In order to demonstrate some important results regarding the validity and 
scope of the previous manipulations, we will consider in this section the 
most usual case in which $m > n$, that is, the number of {\em per capita\/} 
rates is greater than that of variables. For example, in the peroxidase 
equations (16) we have $m=7$ and $n=3$. We will also assume that the rank of 
matrix {\bf B} is maximum: rank({\bf B}) $=n$. 

A necessary condition for ensuring the equivalence between system (9) 
and the generalized Lotka-Volterra equations (13) is that the transformation 
relating them preserves the topological characteristics of the solutions. We 
shall prove that this is indeed the case here. A sufficient condition for 
demonstrating this statement$^{27}$ is the existence of a continuous, 
differentiable and invertible application connecting the initial and final 
phase spaces. Since the dimension of the generalized Lotka-Volterra system is greater 
than that of (9), such an application should connect the phase space of (9) 
and the $n$-dimensional subset of ${\cal R}^{m}$ into which it is mapped.

We can write the transformation (10) relating the power-law rates system 
variables to the {\em per capita\/} rates as:
$$    
    Y_j = \prod_{k=1}^{m} X_{k }^{\tilde{B}_{jk}} ,
      \; \: j = 1, \ldots ,m \; ,      \eqno{(18)}
$$    
where $X_{n+1} = \ldots = X_m = 1$ and $\tilde{{\bf B}}$ is an $m \times m$ 
matrix, defined as:
$$    
   \tilde{{\bf B}} = \left( \begin{array}{c} 
               \mbox{} \\  {\bf B} \\ \mbox{} \end{array} \right. 
     \left| \begin{array}{ccc}
               B'_{1,n+1} & \ldots   & B'_{1,m}  \\
               \vdots     & \mbox{}  & \vdots    \\
               B'_{m,n+1} & \ldots   & B'_{m,m}
     \end{array} \right) \; \equiv \; ( {\bf B} \mid {\bf B'} )   \eqno{(19)}
$$    
Here {\bf B} is the already known $m \times n$ matrix of exponents of the 
{\em per capita\/} rates, and ${\bf B'}$ is a $m \times (m-n)$ matrix of 
arbitrary entries. These entries can always be selected in such a way that 
$\tilde{{\bf B}}$ is invertible. Equation (18) is obviously differentiable. 
Thus, we only need to prove that it is one to one and invertible. If we take 
logarithms in both sides of (18):
$$    
   \left( \begin{array}{c} \ln (Y_{1}) \\ \vdots \\ \ln (Y_{m}) \end{array}
   \right) = \tilde{{\bf B}} \left( \begin{array}{c} \ln (X_{1}) \\ \vdots \\
   \ln (X_{n}) \\ 0 \\ \vdots \\ 0 \end{array} \right)    \eqno{(20)}
$$    
Since rank($\tilde{{\bf B}}$)$=m$, for any two vectors ${\bf X}$ and 
${\bf X'}$ we have $\tilde{{\bf B}} \ln ({\bf X}) \neq \tilde{{\bf B}} 
\ln ({\bf X'})$, unless ${\bf X} = {\bf X'}$. Thus the map (18) is one to 
one and invertible ($\tilde{{\bf B}}$ invertible) and the topology is 
preserved by the transformation.

The original variables of the power-law rates system (9) can be retrieved 
from those of the generalized Lotka-Volterra system by means of two different 
procedures. The first one is obtained by writing system (9) in the separable 
form:
$$    
    \frac{ \dot{X}_i }{X_i} = \sum_{j=1}^{m} A_{ij} Y_j (t) \; , \;\:\;\:\;\: 
    \;\:\;\: i = 1, \ldots , n                    \eqno{(21)}
$$    
Then the $X_i$ result from the formal integrations:
$$    
    X_i (t) = X_i (0) \exp \left\{ \sum_{j=1}^{m} \; A_{ij} \int_{0}^{t} 
    Y_j(t') \mbox{d}t' \right\}                    \eqno{(22)}
$$    
The second approach is purely algebraic and does not require any integration. 
Since {\bf B} is an $m \times n$ matrix, with $m > n$, and rank({\bf B}) is 
maximum, then there exists an $n \times n$ invertible submatrix ${\bf B}_n$ 
of {\bf B}. Let 
$$    
   {\bf B}_n = \left( \begin{array}{ccc}
           B_{i_1 1} & \ldots & B_{i_1 n}  \\
            \vdots   &        & \vdots     \\
           B_{i_n 1} & \ldots & B_{i_n n}  
   \end{array} \right) \; , \;\;\:\;\;\: \{ i_1 , \ldots , i_n \} \subset 
   \{2, \ldots , m \}               \eqno{(23)}
$$    
This implies that:
$$    
   \left( \begin{array}{c} \ln (Y_{i_1}) \\ \vdots \\ \ln (Y_{i_n}) 
   \end{array} \right) = {\bf B}_n \left( \begin{array}{c} 
   \ln (X_{1}) \\ \vdots \\ \ln (X_{n}) \end{array} \right)     \eqno{(24)}
$$    
Since ${\bf B}_n$ is invertible, this finally leads to:
$$    
    X_k (t) = \prod_{p=1}^{n} \left[ Y_{i_p} (t) \right]^{ 
    \left( {\bf B}_n^{-1} \right) _{kp} } \; , \;\:\;\:\;\:\;\: k = 1, 
    \ldots , n \eqno{(25)}
$$    
The time evolution or stability properties of a given reaction network might 
be analyzed in anyone of these two alternative descriptions ({\bf X} or {\bf 
Y}), for they are completely equivalent. However, as far as structural 
properties are concerned, the generalized Lotka-Volterra form (in terms of 
{\bf Y}) seems preferable for it is mathematically characterized by a single 
algebraic object: matrix {\bf L}. 

As we have seen, to every power-law rates system (9) a single generalized 
Lotka-Volterra conjugate system can be associated. The question now is to 
what extent this is also valid for general rate-laws:
$$    
    \dot{{\bf X}} = {\bf F(X)}          \eqno{(26)}
$$    
The way for finding the conjugate representation consists, as we saw in 
Sections II and III in the peroxidase-oxidase example, in the introduction of 
auxiliary variables for functional rate-laws in the right hand side of (26) 
not complying to the power-law rates system format. This always leads to a 
power-law rates system from which the obtainment of the generalized 
Lotka-Volterra system is straightforward. We shall see that, to a great 
extent, the generalized Lotka-Volterra representative is unique for every 
system of the form (26), and is independent of the specific choice of 
auxiliary variables. Instead of a formal approach, we shall consider in more 
detail the peroxidase example. However, the results that we shall display can 
be proved rigorously.$^{22,28}$ 

Let us generalize the procedure of Section II by introducing an auxiliary 
variable of the form: 
$$    
    Z = X^{ \alpha } Y^{ \beta } (1+qX^2)^{ \gamma } \;\:\; ,  \eqno{(27)}
$$    
where $\alpha$, $\beta$ and $\gamma$ are real parameters and $\gamma \neq 0$. 
After some algebra, the introduction of this general variable leads to a  
family of $( \alpha , \beta , \gamma )$-dependent power-law rates systems 
with matrices: 
$$    
{\bf A}(\alpha , \beta , \gamma ) = 
   \left(   \begin{array}{ccccccc}
       -1   & B        &  -  1    &    0    &      0            &       0    &     0     \\
        0   & 0        &     0    &    A    &     -1            &       0    &     0     \\
   - \alpha & B \alpha & - \alpha & A \beta & 2qB \gamma -\beta & -2q \gamma & -2q \gamma 
   \end{array} \right)   \eqno{(28)}
$$    
$$    
{\bf B} (\alpha , \beta , \gamma ) = 
   \left(   \begin{array}{ccc}
           0           &         0             &        0       \\ 
          -1           &         0             &        0       \\ 
   \alpha / \gamma     & 1 + \beta / \gamma    &  -1 / \gamma   \\ 
           0           &        -1             &        0       \\ 
 1 + \alpha / \gamma   &     \beta / \gamma    &  -1 / \gamma   \\ 
 2 + \alpha / \gamma   &     \beta / \gamma    &  -1 / \gamma   \\ 
 2 + 2 \alpha / \gamma & 1 + 2 \beta / \gamma  &  -2 / \gamma   \\ 
   \end{array} \right)    \eqno{(29)}
$$    
However, the product $ {\bf L} = {\bf B} (\alpha , \beta , \gamma ) \cdot 
{\bf A}(\alpha , \beta , \gamma ) $ is independent of $ \alpha $, $ \beta $ 
and $ \gamma $. Since {\bf L} is the matrix associated to the conjugate 
generalized Lotka-Volterra system, this means that such representation is 
unique, independently of the choice of the auxiliary variables. Of course, 
this matrix {\bf L} coincides with the one obtained from the product of (15) 
and (14), which are particular cases of (29) and (28), respectively, with 
$ \alpha = 0, \beta = 0, \gamma = -1 $. This is consistent with the fact that 
the variables of the generalized Lotka-Volterra representative are 
independent of $( \alpha , \beta , \gamma )$. From matrix (29), they are:
\[
   1 \; ; \;\:\; 
   X^{-1} \; ; \;\:\;
   X^{\alpha / \gamma } Y^{ 1 + \beta / \gamma } Z^{ -1 / \gamma } = 
   \frac{Y}{1+ qX^2}  \; ; \;\:\; 
\]
\[   
   Y^{-1}  \; ; \;\:\; 
   X^{1 + \alpha / \gamma } Y^{ \beta / \gamma } Z^{ -1 / \gamma } = 
   \frac{X}{1+ qX^2}  \; ; \;\:\; 
   X^{2 + \alpha / \gamma } Y^{ \beta / \gamma } Z^{ -1 / \gamma } = 
   \frac{X^2}{1+ qX^2}  \; ; \;\:\; 
\]
$$    
   X^{2 + 2 \alpha / \gamma } Y^{ 1 + 2 \beta / \gamma } Z^{ -2 / \gamma } = 
   \frac{X^2 Y}{(1+ qX^2)^2 }          \eqno{(30)}
$$    
This implies that the initial conditions of the conjugate generalized 
Lotka-Volterra system will also be unique. In other words: {\em To every 
general system of the form (26) a single generalized Lotka-Volterra system 
can be associated by means of this procedure.\/} Although the process leads 
to an infinite family of intermediate power-law rates systems, all of them 
possess the same generalized Lotka-Volterra representative, irrespective of 
the parameters $\alpha$, $\beta$ and $\gamma$, and are thus all equivalent. 
This property supports our assertions in favor of the generalized 
Lotka-Volterra as a unifying format.

\pagebreak
\begin{flushleft}
{\bf V. Conclusions}
\end{flushleft}

We have stressed how the power-law formalism can be a referential format for 
general functional forms for chemical rate-laws. On encapsulating a chemical 
mechanism (or network) under a power-law formalism, there is no need to 
resort, as we have shown, to any kind of local approximation in terms of that 
same power-law formalism, even if it seems justified experimentally. Instead, 
simple manipulations of elementary calculus convert non polynomial kinetic 
equations into power-law differential equations, completely equivalent to the 
original ones when appropriate initial conditions are assumed. 

For power-law rates an interesting universal relationship has been obtained. 
When these rate laws are considered as {\em per capita\/} rates (or, 
equivalently, in terms of logarithmic derivatives) they obey a set of 
generalized Lotka-Volterra equations. The specific matrix characterizing this 
generalized Lotka-Volterra system is independent of the particular embedding 
procedure when transforming general rate laws into a power-law formalism. 
Also, to each particular power-law rates system corresponds a unique and 
mathematically equivalent generalized Lotka-Volterra system. The latter may 
then be considered a conjugate representation of any chemical network. 

Much attention has been devoted in the literature to the generalized 
Lotka-Volterra equations, a fact which is not independent of their structural 
simplicity and their ubiquity in many scientific disciplines, ranging from 
population biology to laser physics.$^{26}$ This is particularly attractive 
in the context of classification of chemical networks. 

A {\em conjugate chemical network\/}, with at most bimolecular steps in the 
essential species, may be associated to the generalized Lotka-Volterra 
equations. The network is purely conceptual and should not be thought of as 
the {\em canonical \/} reactional scheme of any chemical process. It should 
be regarded as an abstract equivalent representation of a model system in 
the familiar language of mass-action kinetics. Its inmediate interest in the 
modeling of actual chemical systems may be presently a subject of debate, 
for many critics argue that the field of {\em chemical network dynamics \/} 
has not yet produced any result of chemical importance. This point of view 
should be seriously reconsidered in the light of recent work by Ross and 
collaborators.$^{11}$ 

\mbox{}

\mbox{}

{\bf Acknowledgements:} This work has been supported by the DGICYT \linebreak 
(Spain), under grant PB94-0390. B. H. acknowledges a doctoral fellowship from 
Comunidad Aut\'{o}noma de Madrid. 

\mbox{}

\mbox{}

\begin{flushleft}
{\bf References and notes}
\end{flushleft}
$^1$ \'{E}rdi, P.; T\'oth, J. {\em Mathematical Models of
Chemical Reactions;\/} Manchester University Press: Manchester, 1989; 
pp. 1-13.   \newline
$^2$ Pielou, E. C. {\em Mathematical Ecology;\/} John Wiley \& Sons: 
New York, 1977.   \newline
$^3$ Weidlich, W.; Haag, G. {\em Concepts and Models of a Quantitative 
Sociology;\/} Springer-Verlag: Berlin, 1983.   \newline
$^4$ Hofbauer, J.; Sigmund, K. {\em The Theory of Evolution and 
Dynamical Systems;\/} Cambridge University Press: Cambridge, 1988.  \newline 
$^5$ Murray, J.D. {\em Mathematical Biology,\/} 2nd ed.; Springer-Verlag: 
Berlin, 1993.     \newline 
$^6$ See the articles by Othmer, H.G. and Feinberg, M. in {\em Modelling of 
Chemical Reaction Systems;\/} Ebert, K.H., Deuflhard, P., J\"ager W., Eds.; 
Springer-Verlag: Berlin, 1981.    \newline 
$^7$ Feinberg, M. {\em Arch. Ratl. Mech. Anal.\/} {\bf 1972}, 
{\em 46\/},  1.    \newline 
$^8$ Horn, F.; Jackson, R. {\em Arch. Ratl. Mech. Anal.\/} 
{\bf 1972}, {\em 47\/}, 81.   \newline 
$^9$ See reference 1, pp. 45-48.   \newline 
$^{10}$ Clarke, B. L. {\em Adv. Chem. Phys.\/} {\bf 1980}, {\em 43\/}, 1.  \newline 
$^{11}$ Eiswirth, M.; Freund, A.; Ross, J. {\em Adv. Chem. Phys.\/} 
{\bf 1991}, {\em 80\/}, 127.    
Chevalier, T.; Schreiber, I.; Ross, J. {\em J. Phys. Chem.\/} {\bf 1993}, 
{\em 97\/}, 6776. 
Hung, Y. F.; Ross, J. {\em J. Phys. Chem.\/} {\bf 1995}, {\em 99\/}, 1974. 
Hung, Y. F.; Schreiber, I.; Ross, J. {\em J. Phys. Chem.\/} {\bf 1995}, 
{\em 99\/}, 80. 
Stemwedel, J. D.; Ross, J. {\em J. Phys. Chem.\/} {\bf 1995}, {\em 99\/}, 
1988.    \newline 
$^{12}$ Argyrakis, P. In {\em Fractals, Quasicrystals, Chaos, Knots and 
Algrebraic Quantum Mechanics;\/} Amann, A., Cederbaum, L., Gans, W., Eds.; 
Kluwer: New York, 1988; p. 53.  \newline 
$^{13}$ Klymko, P. W.; Kopelman, R. {\em J. Phys. Chem.\/} {\bf 1982}, 
{\em 86\/}, 3686.
Klymko, P. W.; Kopelman, R. {\em J. Phys. Chem.\/} {\bf 1983}, {\em 87\/}, 
4565.
Anacker, L. W.; Kopelman, R. {\em J. Chem. Phys.\/} {\bf 1984}, {\em 81\/}, 
6402.
Kopelman, R. {\em J. Stat. Phys.\/} {\bf 1986}, {\em 42\/}, 185.
Lin, A.; Kopelman, R.; Argyrakis, P. {\em Phys. Rev. E\/} {\bf 1996}, 
{\em 53\/}, 1502.    \newline 
$^{14}$ See Anacker and Kopelman in reference 13.    \newline 
$^{15}$ Savageau, M. A. {\em J. Theor. Biol.\/} {\bf 1969}, {\em 25\/}, 
365.    \newline 
$^{16}$ Voit, E. O., Ed. {\em Canonical Nonlinear Modelling: S-system 
Approach to Understanding Complexity;\/} Van Nostrand: New York, 1991. \newline 
$^{17}$ See reference 1, p. 64.  \newline 
$^{18}$ Samardzija, N.; Greller, L. D.; Wasserman, E. {\em J. Chem. 
Phys.\/} {\bf 1989}, {\em 90\/}, 2296.   \newline 
$^{19}$ Poland, D. {\em Physica D\/} {\bf 1993}, {\em 65\/}, 86.  \newline 
$^{20}$ Kowalski, K. {\em Chem. Phys. Lett.\/} {\bf 1993}, {\em 209\/}, 
167.   \newline 
$^{21}$ Kerner, E. H. {\em J. Math. Phys. \/} {\bf 1981}, {\em 22\/}, 1366. \newline 
$^{22}$ Hern\'{a}ndez--Bermejo, B.; Fair\'{e}n, V. {\em Phys. Lett. A \/} 
{\bf 1995}, {\em 206\/}, 31. \newline 
$^{23}$ Degn, H.; Harrison, D. E. F. {\em J. Theoret. Biol. \/} {\bf 1969}, 
{\em 22\/}, 238.  \newline 
$^{24}$ Fair\'{e}n, V.; Velarde, M. G. {\em J. Math. Biol. \/} {\bf 1979}, 
{\em 8\/}, 147.   \newline  
$^{25}$ Takeuchi, Y.; Adachi, N.; Tokumaru, H. {\em Math. Biosci. \/} 
{\bf 1978}, {\em 42\/}, 119.   \newline
$^{26}$ Peschel, M.; Mende, W. {\em The Predator--Prey Model. Do
we live in a Volterra World?\/} Springer-Verlag: Wien--New York, 1986. \newline 
$^{27}$ Jackson, E. A. {\em Perspectives of Nonlinear Dynamics,\/} Vol. 1; 
1st ed.; Cambridge University Press: Cambridge, 1994; pp. 21-23. \newline 
$^{28}$ Hern\'{a}ndez--Bermejo, B.; Fair\'{e}n, V. {\em Math. Biosci. \/} 
(in press).
\end{document}